\documentclass[fleqn,twoside]{article}

\usepackage{espcrc2}
\usepackage{subfigure}
\usepackage{graphicx}
\newcommand{\AmS}{{\protect\the\textfont2
  A\kern-.1667em\lower.5ex\hbox{M}\kern-.125emS}}
\title{Multi--Layer Structure  in the Strongly Coupled 5D Abelian Higgs Model}

\author{P. Dimopoulos\address[MCSD]{Physics Department, National Technical University\\
15780 Zografou Campus, Athens, Greece}%
        \thanks{E-mail: pdimop@central.ntua.gr,kfarakos@central.ntua.gr,
        Stam.Nicolis@phys.univ-tours.fr},
        K. Farakos\addressmark,
        S. Nicolis\address{CNRS-Laboratoire de Math\'ematiques et Physique
Th\'eorique (UMR 6083)
Universit\'e de Tours, Parc Grandmont, 37200 Tours,France}}

\begin{document}

\begin{abstract}
We  explore the phase diagram of the 5--D
anisotropic Abelian Higgs model by
Monte Carlo simulations.
In particular, we study the transition between the confining phase
and the  four dimensional layered Higgs phase. We find that,
in a certain region of the lattice parameter space, this
transition can be first order and that each layer  moves into
the Higgs phase independently of the others (
decoupling of layers). For more information about the layer phase
formation see (\cite{funiel}--\cite{dim2}) and the talk by S.
Nicolis \cite{stam2} in the same volume.
\end{abstract}

\maketitle

\section{Formulation of the model}

We write down the lattice action of the 5-D Abelian Higgs model in
typical notation.
Direction $\hat 5$ (transverse)  will be singled
out by couplings (primed) that will differ from the corresponding ones
in the remaining four directions.
%\begin{eqnarray}
%S= &\beta_{g}& \sum_x\sum_{1 \le \mu<\nu \le 4}(1-\cos F_{\mu \nu}(x))
%+\beta_g^{\prime}\sum_x\sum_{1 \le \mu \le 4}(1-\cos F_{\mu 5}(x))
%\nonumber \\
%&+&\beta_{h}\sum _{x} {\rm Re} [4 \varphi^{*}(x)\varphi (x)
%- \sum_{1 \le \mu \le 4} \varphi^{*}(x)U_{\hat \mu}(x) \varphi (x+\hat \mu)]
%\nonumber \\
%&+&\beta_{h}^{\prime} \sum _{x} {\rm Re} [\varphi^{*}(x)\varphi (x)
%- \varphi^{*}(x)U_{\hat 5}(x) \varphi (x+\hat 5)]
%\nonumber \\
%&+&\sum _{x}[(1-2\beta_{R}-4 \beta_{h}- \beta_{h}^{\prime})\varphi^{*}(x)\varphi (x)
%+\beta_{R}(\varphi ^*(x)\varphi (x))^2],
%\label{compactaction}
%\end{eqnarray}
$$
\hspace*{-2.1cm}S= \beta_{g} \sum_x\sum_{1 \le \mu<\nu \le 4}(1-\cos F_{\mu
\nu}(x))$$
$$\hspace*{-1.7cm}+\beta_g^{\prime}\sum_x\sum_{1 \le \mu \le 4}(1-\cos F_{\mu
5}(x))$$
$$+\beta_{h}\sum _{x} {\rm Re} [4 \varphi^{*}(x)\varphi (x)
- \sum_{1 \le \mu \le 4} \varphi^{*}(x)U_{\hat \mu}(x) \varphi (x+\hat
\mu)]$$
$$\hspace*{0.6cm} +\beta_{h}^{\prime} \sum _{x} {\rm Re} [\varphi^{*}(x)\varphi (x)
- \varphi^{*}(x)U_{\hat 5}(x) \varphi (x+\hat 5)]$$
$$\hspace*{-0.6cm}
+\sum _{x}[(1-2\beta_{R}-4 \beta_{h}- \beta_{h}^{\prime})\varphi^{*}(x)\varphi
(x)$$
\begin{equation}
\hspace*{0.6cm} +\beta_{R}(\varphi ^*(x)\varphi (x))^2]
\label{compactaction}
\end{equation}
The order parameters that we use are the plaquette and the link defined
either on 4--dimensional volume or in the transverse direction.
We define also the Higgs measure $R^{2}$ on the 5--D volume. Here,
we show only the space--like link $L_{S}$ and transverse--like
link $L_{T}$:

%\be
%{\rm Space-like~Plaquette:~~~} P_S \equiv <\f{1}{6 N^5} \sum_x
%\sum_{1 \le \mu<\nu \le 4} \cos F_{\mu \nu}(x)>
%\ee
%\be
%{\rm Transverse-like~Plaquette:~~~} P_T \equiv <\f{1}{4 N^5}
%\sum_x \sum_{1 \le \mu \le 4} \cos F_{\mu 5}(x)>
%\ee

$$ \hspace*{-1.7cm} L_S \equiv <\frac{1}{4 N^5} \sum_x
\sum_{1 \le \mu \le 4} \cos(\chi(x+\hat \mu) +$$
\begin{equation}
\hspace*{4cm} A_{\hat \mu}(x)-\chi(x)) >
\end{equation}
\begin{equation}
L_T \equiv <\frac{1}{N^5} \sum_x \cos(\chi(x+\hat 5) + A_{\hat
5}(x)-\chi(x))>
\end{equation}
%\be
%{\rm Higgs~field~measure~squared:~~~} R^2 \equiv \f{1}{N^5} \sum_x \rho^2(x)
%\ee
$N$ is the linear dimension of a symmetric $N^5$
lattice.

\section{Monte Carlo Results \\ and the Confining--Layered Transition}
For the simulations we use 5-hit Metropolis algotithm for the updating
of both the gauge and Higgs fields. In order to get better behaviour
we use a global radial algorithm and an overrelaxation
algorithm for the updating of the Higgs field.
We simulated the system for $4^{5}, 6^{5}, \mbox{and} \hspace{0.1cm} 8^{5}$
lattices. We made use, mainly, of the hysteresis loop technique
to establish the phase diagram of the system.
For the results shown here we set  $\beta_{g}=0.5$,
$\beta^{\prime}_{h}=0.001$ and $\beta_{R}=0.1$.
Thus the phase diagram has been found in the  $\beta_{g}^{\prime} - \beta
_{h}$ subspace. In Figure 1 the  phase diagram for $\beta_{R}=0.1$ is
given. S denotes the strong phase, C is the 5-D Coulomb phase,
$H_{5}$ is the 5-D Higgs phase and $H_{4}$ is the layer Higgs phase in
4-D. Further analysis for the phase diagrams
and for the order of the phase transitions can be found in
\cite{dimo3}.

\begin{figure}[!h]
\begin{center}
\includegraphics[scale=0.30]{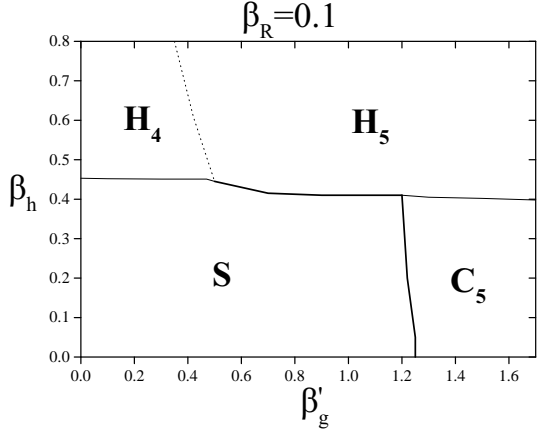}
\caption{Phase diagram for $\beta_{R}=0.1$}\label{Fig2}
\end{center}
\end{figure}

\begin{figure}[!h]
\begin{center}
\subfigure[]{\includegraphics[scale=0.24]{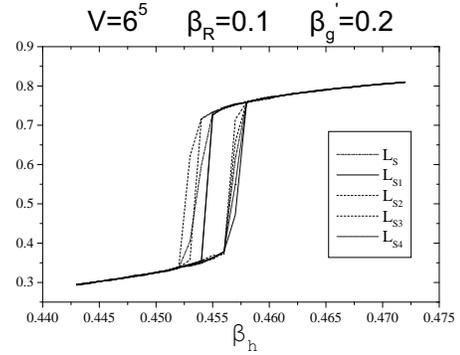}}
\subfigure[]{\includegraphics[scale=0.23]{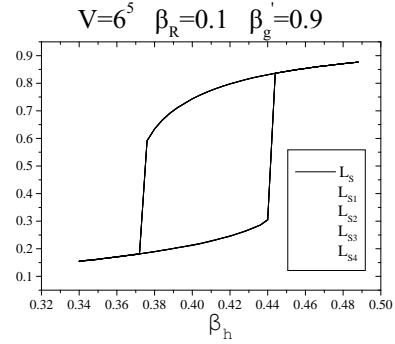}}
\caption{Hysteresis loop for $\beta_{g}^{\prime}=0.2$ (a) and
$\beta_{g}^{\prime}=0.9$ (b) measuring the order parameter on
every layer.} \label{Fig11}
\end{center}
\end{figure}

\begin{figure}[!h]
\subfigure[]{\includegraphics[scale=0.23]{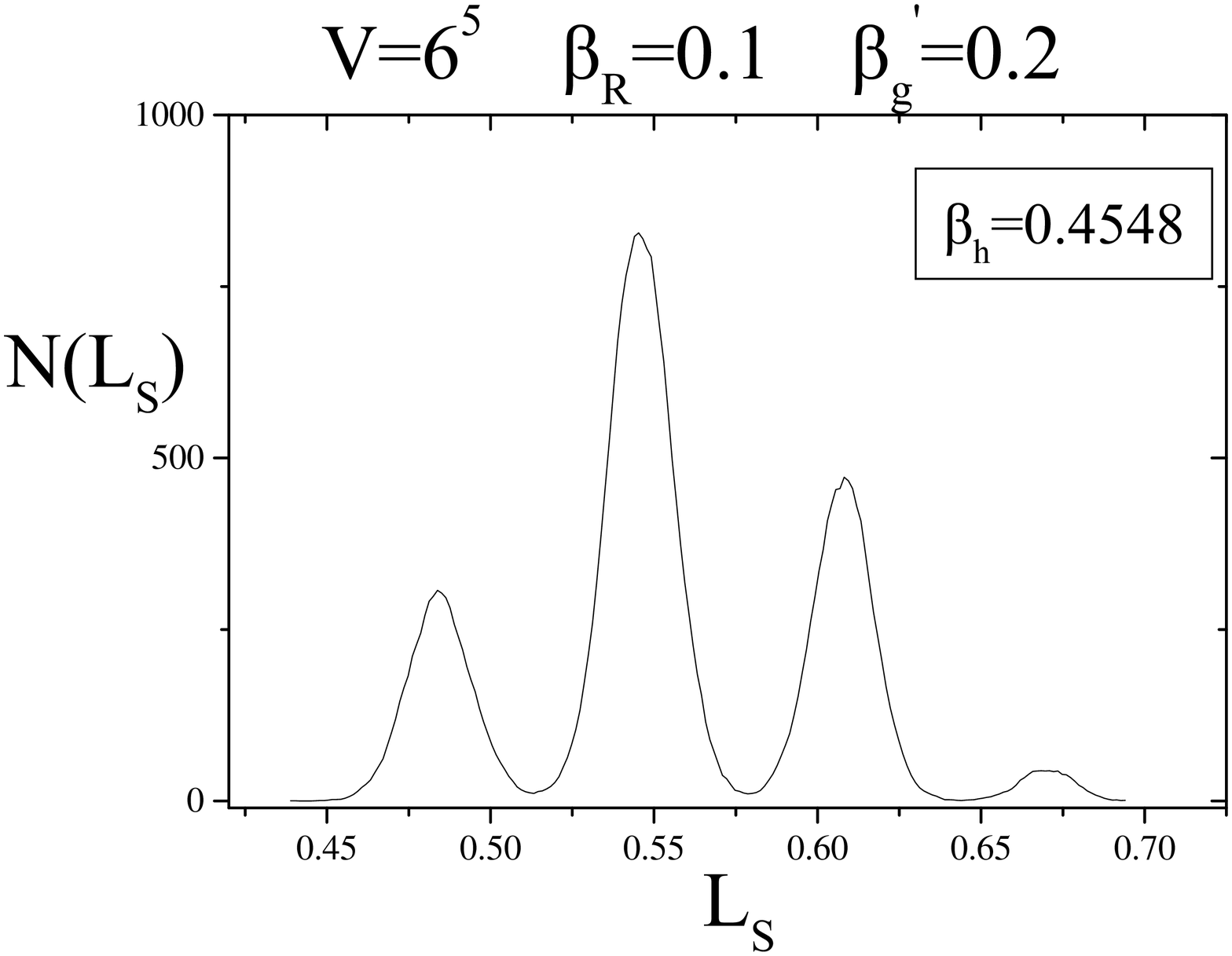}}
\subfigure[]{\includegraphics[scale=0.23]{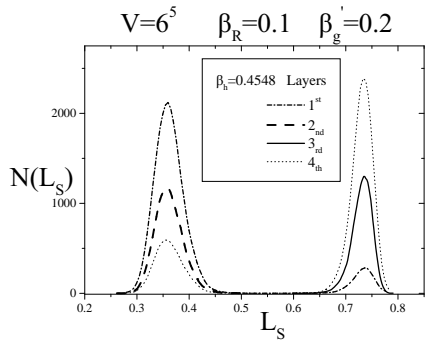}}
\caption{Distribution for $L_{S}$ on the volume (a) and on each
layer (only four of the six are shown in the figure)(b) for $6^5$
lattice  in the critical region for the $S - H_{4}$ phase
transition.} \label{Fig13}
\end{figure}

%\begin{figure}[!h]
%\begin{center}
%\subfigure[]{\includegraphics[scale=0.2]{hpsSHbr01.eps}}
%\subfigure[]{\includegraphics[scale=0.2]{hptSHbr01.eps}}
%\caption{Hysteresis loop for
%space--like (a) and transverse--like (b) plaquette
%for two values of $\beta_{g}^{\prime}=0.2$ and 0.7.}
%\label{Fig3}
%\end{center}
%\end{figure}

\subsection{The Multi--Layer structure}
In Figure 2 we show the very different way
for the transition of $L_{S}$ calculated either on the
five dimensional volume and on four layers. In the figure 2a
 the $S - H_{4}$ transition is presented where the layers show
a {\bf decoherent} behaviour on the phase transition in contrast with
the $S - H_{5}$ phase transition Fig. 2b where the transition
for $L_{S}$ is identical for all layers, so it coincides with the
mean values over the 5D--volume. This specific behaviour of hysteresis loops, may
actually  serve as  a ``criterion" to characterise
the layered phase.

This  behaviour can be  further confirmed by long runs
in the transition region. From the hysteresis loop results found
it is to be expected
that the $S - H_{4}$ phase transition is of first order. Therefore,
we would expect a two peak signal in the order parameter
disrtibution at equilibrium. Nevertheless the situation, shown in
Figure 3(a) concerning the distribution for the order parameter
$L_{S}$  for a value of $\beta_{h}$
near  the transition region  for $V=6^5$
is by no means what one would expect normally. The multipeak structure
seems rather strange. {It should be noticed that the same occurs for
the order parameter $P_{S}$ too. }However, the study of the same order
parameter defined on each space--like volume is more illuminating. For
example in Figure 3(b), we can see
the distribution of $L_{S}$
values on each layer.
We show four out of six distributions of $L_{S}$ corresponding to the
four  space--like layers within the five--dimensional volume.
The  distributions which are produced appear quite usual and they show
that at the same time one layer is in the strong phase (called $2_{nd}$ in
the figure), another has already
passed to the broken phase ($3_{rd}$) and others produce a two peak signal result.
The result is that the strange picture of the distribution formed in Figure
3(a) is  resolved if we analyze the behaviour of
the system on each layer as the system undergoes the phase
transition.
This result is found  for all of the  volume sizes (e.g. $4^5, 6^5, 8^5$)
which we have worked on. Although it is consistent with a first
order phase transition it lends support to the view of
a  {\it decoherent} behaviour for every four--dimensional
volume (layer) in the transition region between five--dimensonal
strong phase and the four--dimensional layered Higgs phase.
This certain behaviour describes a dynamical decoupling of the
layers and provides a possible mechanism for localization of
the fields on the layers.

\section{Conclusions}
We have shown,
by using Monte--Carlo methods,  that a layered Higgs  phase actually exists
in the phase diagram of the five--dimensional Abelian
Higgs model. Furthermore, this layered phase is separated by a phase
transition  with the five--dimensional strong phase.
In the region of  parameter space that we explored
the phase transition in question is first order and
its additional feature is the multi--layer formation. The latter
occurs  as a consequence of the independent formation of
each layer  as the phase transition takes place.

{\bf Acknowledgements}: The authors are grateful to G. Koutsoumbas and A. Kehagias for useful discussions.
The research of P.D. and K.F. has been partially supported by
the TMR Network ``Finite Temperature phase transitions in Particle Physics",
EU contract number FMRX-CT97-0122.

\end{document}